\documentclass{appolb}
\usepackage{cite,epsfig}

\newcommand{\be}{\begin{equation}}
\newcommand{\ee}{\end{equation}}
\newcommand{\ba}{\begin{eqnarray}}
\newcommand{\ea}{\end{eqnarray}}

\begin{document}

\begin{titlepage}
\begin{flushright}
CAFPE-121/09\\
UF-FT-251/09\\
\end{flushright}
\vspace{2cm}
\begin{center}

{\large\bf Standard Model Prediction of the Muon Anomalous Magnetic Moment
\footnote{Invited talk given by J.P.
at Topical FLAVIAnet Workshop on ``Low Energy Constraints on Extensions 
of the Standard Model'', July 24-26  2009, Kazimierz, Poland.}}\\
\vfill
{\bf  Joaquim Prades}
\\[0.5cm]

 CAFPE and Departamento de
 F\'{\i}sica Te\'orica y del Cosmos, Universidad de Granada, 
Campus de Fuente Nueva, E-18002 Granada, Spain.

\end{center}
\vfill
\begin{abstract}
\noindent
I review the present Standard Model  prediction
of the muon anomalous magnetic moment.
The discrepancy with its experimental determination 
is   $(25.5 \pm 8.0) \times 10^{-10}$,
i.e., 3.2 standard deviations. 
\end{abstract}
\vfill
September  2009
\end{titlepage}
\setcounter{page}{1}
\setcounter{footnote}{0}





\section{Introduction}

 The general vertex $\Gamma_\mu$  between a 
fermion $f$  and  an external electromagnetic (EM) field $A^\mu(q=p-p')$
can be written as
\ba
\overline{u}(p') \Gamma_\mu u(p) =
\overline{u}(p') \left[ \gamma_\mu \, F_1(q^2) +
i \, \frac{\sigma_{\mu\nu} q^\nu}{2 m_f} \, F(q^2) + 
\cdots \right]
\ea
 The magnetic dipole moment for a charged
fermion ($f=e, \mu, \tau , \cdots$) is proportional to the spin
$\vec{s}$ through the gyromagnetic 
factor $g_f\equiv 2 (F_1(0)+F_2(0))$
\ba
\vec{\mu} = g_f \frac{e}{2 m_f} \, \vec{s} \, .
\ea
 The Dirac vertex predicts $F_1(0)=1$ and $F_2(0)=0$ at tree-level, 
quantum loops modify this prediction to $F_2(0) \equiv a_f \neq 0$ while
$F_1(0)$ is a conserved charge. The quantity $a_f=(g_f-2)/2$  
is  commonly called the fermion anomaly.

The  muon anomalous magnetic moment $a_\mu$  has been measured by the
E821 experiment (Muon g-2 Collaboration)
at BNL with an impressive accuracy of 0.72 ppm \cite{BNL06}
yielding the present world average
\be
a_\mu^{\rm exp} = (11 ,\, 659 ,\, 208.9 \pm 6.3) \times 10^{-10} \,
\label{amuexp}
\ee
with an accuracy of 0.54 ppm.
 New experiments \cite{fermilab,jparc} are being designed to
measure $a_\mu$ with an accuracy better than 0.14 ppm.
  Here, I'm interested in what is the present status of the
Standard Model (SM) prediction for this very precise measurement.
Is there room for new physics in $a_\mu$?
Recent reviews can be found in 
\cite{EdR09,JN09,MRR07,JEG08,PAS05,KNE04}.

First, I will shortly recall recent advances in the electron
anomaly $a_e$ which is a necessary ingredient to predict
the muon anomaly $a_\mu$. I'll discuss then briefly the main contribution
to $a_e$ that is QED, and the much smaller hadronic and weak contributions.
Secondly, I'll discuss briefly the main contribution
to $a_\mu$ that is QED and the smaller weak contribution.
Then I discuss the also smaller but dominant in the uncertainty
hadronic contribution to $a_\mu$.
Finally, I'll give the conclusions and prospects for improving
$a_\mu$ both theoretic and experimentally.

\section{The electron anomaly $a_e$}

 The Harvard group \cite{HARV08} has very recently made a very precise
measurement of the electron anomaly 
\ba
a_e^{\rm exp} = (11 ,\, 596 ,\, 521 ,\, 807.3  \pm 2.8) \times 10^{-13} 
\ea
 improving the  1987 measurement of the University of Washington group
\cite{WAS87} 
\ba
a_e^{\rm exp} = (11 ,\, 596 ,\, 521 ,\, 883 \pm 42) \times 10^{-13} 
\ea
by more than a factor 15.

 The new $a_e$ measurement provides the most accurate value
for the fine structure constant \footnote{For a very recent discussion 
of the extraction of $\alpha$ from $a_e$ and other methods, see 
\cite{GAB09}.}

\ba
\alpha^{-1} = 137.035 \, 999 \, 084(33)(39) =
137.035 \, 999 \, 084(51) \, 
\label{alpha}
\ea
which is one order of magnitude  more precise than the best previous
determination\footnote{Notice that due to the  2007 corrected value
 for the four-loop order QED coefficient \cite{C8new},
 the value quoted in both PDG 2008 \cite{PDG08} 
and CODATA 2006 \cite{CODATA06} 
 $\alpha^{-1} = 137.035 \, 999 \, 679(94)$ is not correct.}.
 The best non-$a_e$ based determinations of $\alpha$
come from atomic physics, and in
 particular  from the precise measurement of  Cs \cite{Cs02}
and Rb \cite{Rb08} atomic masses,
 \ba
{\rm Cs \,\, atom} \,\,  (2002): 
\alpha^{-1} = 137.036 \, 000 \, 000(1 \, 100) \, \nonumber  \\
{\rm Rb  \,\, atom} \,\, (2008):
\alpha^{-1} = 137.035 \, 999 \, 450(620) \, .
\ea
 It is expected that the Rb atomic mass measurement
can reach the level of accuracy of $a_e$ determinations soon. 
 At present, $a_e$ together from $\alpha$ from
Cs and Rb atomic mass determinations checks QED at 3-loops !
If both determinations were at the same level of accuracy, it would check
QED at 4-loops and a possible substructure of the electron.

\section{Standard Model Contributions to $a_e$}

The dominant SM contribution comes from QED
\ba
a_e^{\rm QED} = \sum_{n=1} \, C_{2n} \, \left(\frac{\alpha}{\pi}
\right)^n \, 
\ea
with
$C_2=1/2$ calculated by Schwinger \cite{SCH48} 61 years ago;
one can find the values of  the rest of the coefficients  
in different reviews \cite{EdR09,JN09,JEG08,MRR07,PAS05,KNE04}.
 The coefficients $C_4$ \cite{C4} and $C_6$ \cite{C6} are known 
analytically including lepton mass  corrections.
$C_8$ is only numerically known  and  being 
continuously improved since early 1980s \cite{C8}. 
Actually, it was corrected recently in 2007 \cite{C8new}. For $C_{10}$,
one uses $0.0 \pm 4.6$ as an estimate where the error
is based on the $C_{2n}$ values growing.
 Its full numerical calculation is in progress, it involves
12,672 diagrams  
and some partial results are already available \cite{C10}.
Recently, specific classes of the QED eighth and tenth orders have been
calculated analytically using Mellin-Barnes transform techniques
\cite{AGdR08} and agree with the numerical results in 
 \cite{C8,C8new,C10}.

There are other much smaller contributions to $a_e$ but that at
 the level of present precision start to be needed. 
 The largest is the leading order hadronic vacuum polarization 
contribution (see Fig 1)
\cite{DH98}
\ba
a_e^{\rm LO \, Hadronic}= (18.75\pm 0.18) \times 10^{-13} \, .
\ea
 while higher order hadronic vacuum polarization contribution is
\cite{KRA97}
\ba
a_e^{\rm HO \, Hadronic}= -(2.25\pm 0.05) \times 10^{-13} \, 
\ea
and the hadronic light-by-light contribution (see Fig 2) is 
\cite{PdRV09}
\ba
a_e^{\rm Hadronic \, LbL}= (0.35\pm 0.10) \times 10^{-13} \, .
\ea
The total hadronic contribution to $a_e$ is
\ba
a_e^{\rm Hadronic}= (16.85\pm 0.21) \times 10^{-13} \, .
\ea
Notice that this is the first time that $a_e^{\exp}$ is sensitive
to the hadronic contribution.
The electroweak contribution is much smaller \cite{JN09}  
\ba
a_e^{\rm EW}= (0.385\pm 0.004) \times 10^{-13} \, .
\ea

\section{Standard Model Contributions to $a_\mu$}

\subsection{QED}

 A precise value of the fine structure constant 
is  very important to the determination of
the muon anomaly since QED is again the dominant
SM contribution. One can write the QED contribution as
\ba
a_\mu^{\rm QED} = \sum_{n=1} \, C_{2n} \, \left(\frac{\alpha}{\pi}
\right)^n \,  
\ea
 and the same comments and authors can be quoted for the
$C_{2n}$ coefficients that can be found in the reviews
\cite{EdR09,JN09,MRR07,JEG08,PAS05,KNE04}. For $C_{10}$, one uses
$663\pm 20$  which is also  estimate and  its calculation is progress
in parallel to  $C_{10}$ for $a_e$ \cite{C10}.
 Using those coefficients and  
$\alpha$ from the latest $a_e$ in (\ref{alpha}), one gets
\ba
a_\mu^{\rm QED} = (11 ,\, 658 ,\, 471.810 \pm 0.015) \times 10^{-10}
\ea
where the largest uncertainty comes from $C_{10}$.
Given that precision, 
the difference between the experimental value (\ref{amuexp}) and its QED
prediction 
\ba
\Delta_{\rm Not \, QED}^{\rm exp} = 
(736.2 \pm 6.3) \times 10^{-10}
\ea
can be considered as a ``measurement''.
In fact, $a_\mu$ has been sensitive  to the hadronic contribution
since 1975 and to the electroweak ones since 2001.

\subsection{Electroweak}

The one loop electroweak contribution is fully known since 1972
\cite{EWone}
\ba
a_\mu^{\rm EW \, one-loop} &=& \frac{5 G_\mu m^2}{24\sqrt{2} \pi^2}
\left[1+ \frac{1}{5}\left(1-4\sin^2(\theta_W) \right)^2 +
{\cal O} \left(m/M_{Z,W,H}\right) \right] \nonumber  \\
&=& (19.5 \pm 0.2) \times 10^{-10} \, . 
\ea

 The full two-loops electroweak contribution has been finished
much more recently (in 2004) though its potential importance was already
 pointed out  in 1992 \cite{KKS92}.  The full result can be found
in \cite{EWtwo}.

The final result for the two-loop electroweak contribution 
\cite{CKM96,EWtwo}
is $- (4.1 \pm 0.1) \times 10^{-10}$  and the total
electroweak contribution 
\ba
a_\mu^{\rm EW} = 
\left[ (19.5 \pm 0.2) -(4.1 \pm 0.1) \right] \times 10^{-10}
= (15.4 \pm 0.2) \times 10^{-10} \, .
\ea
Again, given the experimental precision, 
the difference between the precise  
experimental value in (\ref{amuexp}) and its QED plus EW prediction 
\ba
\Delta_{\rm Not \, QED+EW}^{\rm exp} = 
(720.8 \pm 6.3) \times 10^{-10}
\ea
can be considered as a ``measurement''. This also tells
us that we need to know the hadronic contribution
to $a_\mu$ with an accuracy below  the 1 \% level!

\subsection{Hadronic Contributions}

\subsubsection{Vacuum Polarization Contribution}

 This  contribution  is depicted in Fig. 1.
\begin{figure}
\label{hadrons}
\begin{center}
\epsfig{file=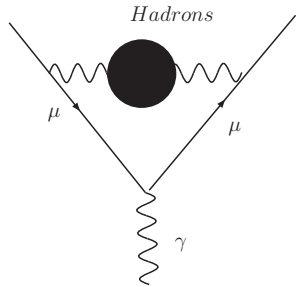,width=4cm}
\end{center}
\caption{Lowest order hadronic 
vacuum polarization contribution to  muon $g-2$.}
\end{figure}
Its leading-order (LO)  contribution is order $\alpha^2$
and can be written in terms of
the one-photon $e^+ e^- \to \gamma^* \to {\rm hadrons}$  
cross-section,  $\sigma^{(0)}(s)$  \cite{LOhad}.
\ba
a_\mu^{\rm LO \, Hadronic} =
\frac{1}{4 \pi^3} \int^\infty_{4 m_\pi^2} \, 
{\rm d}s \, K(s) \, \sigma^{(0)}(s)
\ea
with
\ba
K(s) = \int^1_0 {\rm d}x \, \frac{x^2(1-x)}{x^2+\frac{s}{m^2}(1-x)} \, .
\ea
The kernel $K(s)$ behaves as $m^2/s$ making the low-energy region
($\rho$-region) dominate in the integrand. In fact, the
$\pi^+\pi^-$ channel below 1 GeV gives 72 \% of the total
of  $a_\mu^{\rm LO \, Hadronic}$. 

The one-photon 
cross-section $\sigma^{(0)}(s)$ is (almost) experimentally
obtained. One has to correct the experimental 
cross-section to include final state photons and exclude
the running of $\alpha(s)$  not to make double counting with
higher orders. It is also welcome to eliminate
systematics to have the cross-section $\sigma^{(0)}$
 normalized to  the $e^+e^- \to \mu^+ \mu^-$
cross-section measured by the same experiment.
In the case of the 
initial state radiation (ISR)  method,  this 
is  done by BaBar which measures 
$e^+e^- \to \pi^+ \pi^- \gamma (\gamma)$
 normalized to $e^+e^- \to \mu^+ \mu^- \gamma (\gamma)$
cross-section measured by the same experiment
and is  planed to be done in the forthcoming KLOE-2 measurement.

 For the low energy contribution to $\sigma^{(0)}$, 
we have very precise $e^+e^-$ data from several low-energy
experiments, namely,  from CMD-2 and SND detectors at 
VEPP-2M (Novosibirsk), KLOE at DA$\Phi$NE (Frascati)
and BaBar at PEP-II (SLAC).  The last results from
CMD-2 \cite{CMD207,CMD206} and SND \cite{SND06} are in nice agreement. 
Final KLOE data \cite{KLOE09} are at the same
level of accuracy as CMD-2 and SND data
and there is  an overall agreement at energies between
0.630 and 0.958 GeV with CMD-2 and SND though KLOE data 
lies somewhat lower. Very recently, BaBar using the ISR method
has also released $e^+e^-$ results for the
dominant $\pi^+\pi^-$  at energies between 0.3 GeV and 3 GeV
\cite{BABAR09}.  Belle at KEK (Tsukuba)
which has larger statistics, will do so soon.

At high energy, higher than 13 GeV, and in between 
$\overline c c$ and $\overline b b$ thresholds, perturbative QCD
is used to calculate $\sigma^{(0)}(s)$.

 Combining  all precision $e^+e^-$ data, including the latest
BaBar data \cite{BABAR09}, the authors \cite{DAV09b} quote
\ba
\label{amuLOhad}
a_\mu^{\rm LO \, Hadronic} = (695.5 \pm 4.0_{\rm exp} \pm 0.7_{\rm QCD})
\times 10^{-10} = (695.5 \pm 4.1) \times 10^{-10} \,  .
\nonumber \\
\ea
A new analysis of isospin corrections to 
$\tau^+ \to  \pi^+ \pi^0 \nu$ data
when compared to the  the CVC $e^+e^- \to \pi^+ \pi^-$ data  
has been also recently released \cite{DAV09a}.
The discrepancy between $\tau$ data and the combined $e^+e^-$
data for the dominant $\pi^+\pi^-$ contribution to $\sigma^{(0)}$
of the measurements quoted above
  is reduced to 1.5 $\sigma$ from the previous 4.5 $\sigma$  discrepancy
\cite{DAV09b,DAV09a}.

There are also hadronic vacuum polarization contributions at  
higher order included in Fig 1. The most recent evaluation including
$\alpha^3$ corrections is
 in \cite{HAG07} using also $e^+e^-$ data, 
\ba
a_\mu^{\rm HO \, Hadronic} = -(9.79 \pm 0.08_{\rm exp}
\pm 0.03_{\rm rad}) \times 10^{-10} =
-(9.79 \pm 0.09) \times 10^{-10} \, .
\nonumber
\ea
Where the uncertainty also takes into account 
 non-included radiative corrections \cite{HAG07}.

\subsubsection{Light-by-Light Contribution}

One of the six possible momenta routing 
to the hadronic light-by-light
contribution  to $a_\mu$ is in  Fig. 2.
\begin{figure}
\label{lbl}
\begin{center}
\epsfig{file=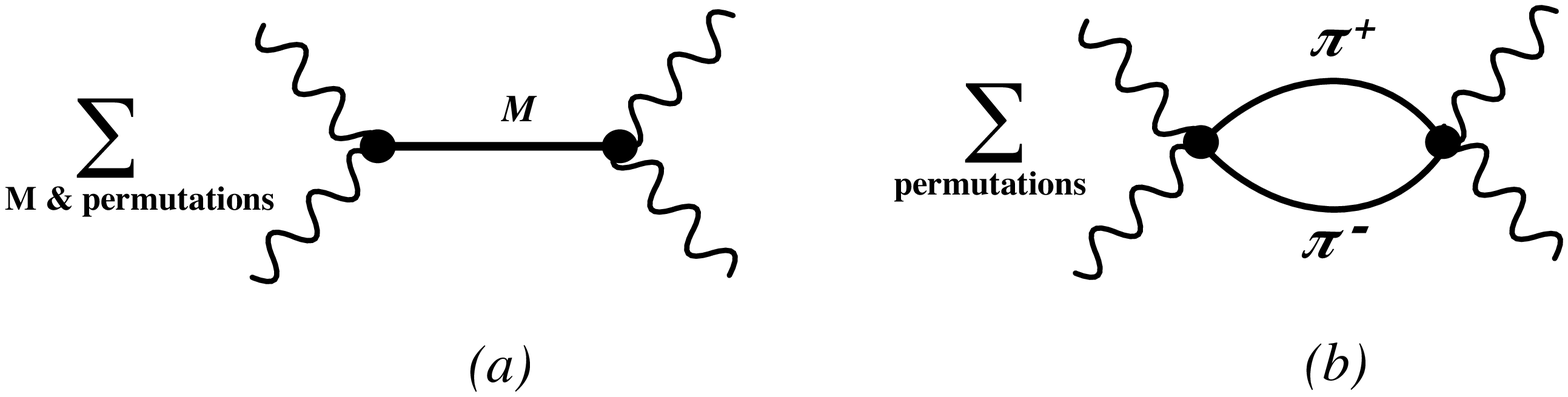,width=4cm}
\end{center}
\caption{Hadronic light-by-light
contribution to $a_\mu$  is in  Fig. 2.}
\end{figure}
 Recent work on   $a^{\rm HLbL}$ can be found in 
\cite{PdRV09,HK98,BPP96,BP01,KN02,MV04,DB08,NYF09,LATT09}. 
For reviews on the status, prospects of $a_\mu^{\rm HLbL}$
and references see \cite{reviews,EdR09,JN09}.
In particular, the  discussion in \cite{PdRV09} 
 lead the authors to give  the following value 
\be
a_\mu^{\rm Hadronic  \, \, LbL} = (10.5 \pm 2.6) \times 10^{-10} \, 
\ee
for their present best estimate.

\section{Result for $a_\mu^{\rm SM}$}

Adding all pieces discussed above contributing to $a_\mu$, one gets
\ba
\label{amuth}
10^{10} a_\mu^{\rm SM} &=& 
\overbrace{(11 ,\, 658 ,\, 471.810 \pm 0.015)}^{\rm{QED}} +
 \overbrace{(15.4 \pm 0.2)}^{\rm{EW}}  \nonumber \\
&+& \underbrace{(695.5 \pm 4.1)}_{\rm{LO \, Had}} 
 \underbrace{-(9.79 \pm 0.09)}_{\rm{HO \, Had}} + 
\underbrace{(10.5 \pm 2.6)}_{\rm{Had \, LbL}} \nonumber
\\ &=&
\underbrace{(11 ,\, 658 ,\, 487.2  \pm 0.2)}_{\rm{QED + EW}} +
\underbrace{(696.2 \pm 4.9)}_{\rm{Hadronic}} \nonumber \\
&=&  11 ,\, 659 ,\, 183.4 \pm 4.9  \, . 
\ea

Using this result and the experimental value of $a_\mu$
in (\ref{amuexp}), one gets
\ba
\label{disc}
a_\mu^{\rm exp} - a_\mu^{\rm SM} = 
(25.5 \pm 8.0) \times 10^{-10} \, .
\ea

\section{Conclusions and Prospects}

Combining all
 recent  precise $e^+e^-$ data obtained with  different methods
(energy scanning and ISR) in different experiments 
to calculate $\sigma^{(0)}$   
\cite{DAV09b} and using the new evaluation of the 
hadronic light-by-light contribution in \cite{PdRV09,reviews} one gets 
more than  3 $\sigma$ of discrepancy between the SM
value for $a_\mu$ (\ref{amuth}) and its experimental determination
(\ref{amuexp}). This discrepancy has been slowly
growing due to impressive theory and experiment recent achievements. 
 In fact,  both theory and experiment  
uncertainties have been reduced by more than a factor two
in the last eight years
\footnote{In 2001 , this discrepancy was 
$a_\mu^{\rm exp} - a_\mu^{\rm SM} = (23.1 \pm 16.9) \times 10^{-10}$.
\cite{PRA01}}.

 There are planned new $e^+e^-$ experiments at Novosibirsk
(VEPP-2000) and Frascati (DA$\Phi$NE-2) which will cross-check
 present results and reduce the present uncertainty in  
$a_\mu^{\rm LO \, Had}$ which dominates the final SM uncertainty now.
There are also new $\tau$ data at B-factories and a new $\tau$-charm 
factory at Beijing which will cross-check the $\tau$ result.
It will also help to understand better the isospin violation corrections
in order to use the forthcoming precise $\tau$ data.
  In addition, a new full calculation
of $a_\mu^{\rm Had \, LbL}$ is desirable and possible. The goal in this
case is to reduce its present uncertainty to the level of 
$(1.5 \sim 2.0) \times 10^{-10}$. With all these eventual
theory improvements the uncertainty of the SM prediction can be
reduced soon enough for the new experiments being designed.
 As said before, these  experiments are being  designed to
measure $a_\mu$ with an accuracy better than 0.14 ppm
in parallel to the  expected theory advances.
 The proposed experiment at Fermilab is designed to reduce
the $a_\mu$ uncertainty to the level of $1.6 \times 10^{-10}$
\cite{fermilab} and the one  at J-PARC to somewhere 
between $1.2 \times 10^{-10}$ and  $0.6 \times 10^{-10}$ \cite{jparc}.

With all the  theory activity detailed above 
and the tantalizing more than 3 $\sigma$ 
discrepancy in $a_\mu$ (\ref{disc})
between the SM prediction  and its experimental determination, 
I believe that those new $g-2$ experiments are very timely,
necessarily complementary to direct searches like LHC and ILC
and should be done as soon as possible. Its very likely that
they give the first new physics discovery ! 

\section*{Acknowledgments}

This work is supported in part by the European Commission (EC) 
RTN network,
Contract No.  MRTN-CT-2006-035482  (FLAVIAnet), by the Spanish 
Consolider-Ingenio 2010 Programme CPAN Grant No. CSD2007-00042,
 MEC (Spain) and  FEDER (EC) Grant No. FPA2006-05294,  Junta
de Andaluc\'{\i}a Grant Nos.  P05-FQM 437 and P07-FQM 03048.


\begin{thebibliography}{0}

\bibitem{BNL06}
B.L. Roberts, arXiV:1001.2898;
G.W. Bennett {\it et al.}  [Muon $g-2$ Collaboration],
  Phys.  Rev.  D {\bf 73} (2006) 072003;
 Phys.\ Rev.\ Lett.\  {\bf 92} (2004) 161802.

\bibitem{fermilab}
D.W. Hertzog, 
Nucl. Phys. B (Proc.\ Suppl.)  
\textbf{181-182} (2008) 5;
 D.W. Hertzog, {\it et al}
  arXiv:0705.4617;
B.L. Roberts  [E821 and E969 Collaborations],
  Nucl.\ Phys.\ B (Proc.\ Suppl.)  {\bf 155} (2006) 372;
  hep-ex/0501012.

\bibitem{jparc}
J. Imazato,
Nucl. Phys. B (Proc. Suppl.) \textbf{129-130} (2004) 81;
``Letter of Intent: An Improved Muon $(g-2)$ Experiment at J-PARC'',
 R.M. Carey {\it et al}. (2003).

\bibitem{EdR09}
E. de Rafael, 
 Nucl.\ Phys. B (Proc.\ Suppl.)  {\bf 186} (2009) 211;
   PoS {\bf(EFT09)} 050.

\bibitem{JN09}  F. Jegerlehner and A. Nyffeler,
  Phys. Rept. \textbf{477} (2009) 1.

\bibitem{JEG08}
  F. Jegerlehner,  Lect.\ Notes Phys.\  \textbf{745} (2008) 9;
  Acta Phys.\ Polon.\  B \textbf{38} (2007) 3021.

\bibitem{MRR07} 
J.P. Miller, E. de Rafael and B.L. Roberts,
  Rept.\ Prog.\ Phys.\  {\bf 70} (2007) 795.

\bibitem{PAS05}
 M. Passera,  J.\ Phys.\ G {\bf 31} (2005) R75.

\bibitem{KNE04}
M. Knecht,
  Lect.\ Notes Phys.\  {\bf 629} (2004) 37.

\bibitem{HARV08} 
D. Hanneke, S. Fogwell, and G. Gabrielse, 
Phys. Rev. Lett. {\bf 100} (2008) 120801;
B. Odom, {\it et al}, ibid. {\bf 97} (2006) 030801.

\bibitem{WAS87}
R.S. Van Dyck, Jr., P.B. Schwinberg, and H.G. Dehmelt, 
Phys. Rev. Lett. {\bf 59} (1987) 26.

\bibitem{GAB09}
G. Gabrielse, 
in ``Lepton Dipole Moments'', B.L. Roberts and W.J. Marciano (eds)  
(World Scientific, Singapore, 2009), 
Advanced Series on Directions in High Energy Physics, Vol. 20.

\bibitem{C8new}
  T. Aoyama {\it et al.}
  Phys.\ Rev.\  D {\bf 77} (2008) 053012.

\bibitem{PDG08}
C. Amsler, {\it et al}, Phys. Lett. B {\bf 667} (2008) 1.

\bibitem{CODATA06}
P.J. Mohr, B.N. Taylor and D.B. Newell,
 Rev. Mod. Phys. 80 (2008) 633.

\bibitem{Cs02}
A. Wicht, {\it et al}, Phys. Scripta {\bf T102} (2002) 82.

\bibitem{Rb08}
M. Cadoret, {\it et al}, Phys. Rev. Lett. {\bf 101} (2008) 230801.

\bibitem{SCH48}
J.S. Schwinger, Phys. Rev. {\bf 73} (1948) 416.

\bibitem{C4}
C.M. Sommerfield, Phys. Rev. {\bf 107} (1957);
 Ann. Phys. (N.Y.)  {\bf 5} (1958) 677;
A. Petermann,  Hel. Phys. Acta {\bf 30} 407;
Nucl. Phys. {\bf 5} (1958) 677;
H. Suura and E. Wichmann, Phys. Rev. {\bf 105} (1957) 1930;
A. Petermann, ibid. {\bf 105} (1957) 1931;
H.H. Elend, Phys. Lett. {\bf 20} (1966) 682;
[Errat-ibid {\bf 21} (1966) 720.]

\bibitem{C6}
J.A. Mignaco and E. Remiddi, Nuovo Cim. A {\bf 60} (1969) 519
R. Barbieri and E. Remiddi, Phys. Lett. B {\bf 49} (1974) 468;
Nucl. Phys. B  {\bf 90} (1975) 233;
R. Barbieri, M. Caffo and E.  Remiddi, Phys. Lett. B {\bf 57} (1975) 460;
M. Levine, E. Remiddi and R. Roskies, Phys. Rev. D  {\bf 20} (1979) 2068;
S. Laporta and E. Remiddi, Phys. Lett. B {\bf 265} (1991) 182;
ibid. {\bf 301} (1993) 440;
ibid.{\bf 356} (1995) 390;
ibid. {\bf 379} (1996) 283;
S. Laporta, Phys. Rev. D {\bf 47} (1993) 4793; 
Phys. Lett. B {\bf 343} (1995) 421; 
Nuovo Cim. A  {\bf 106} (1996) 675;
T. Kinoshita, Nuovo Cim. B {\bf 51} (1967) 140;
B.E. Lautrup and E. de Rafael, Phys. Rev. {\bf 174} (1968) 1835;
B.E. Lautrup and M.A. Samuel, Phys. Lett. B {\bf 72} (1977) 114;
M.A. Samuel and G. Li, Phys. Rev. D {\bf 44} (1991) 3935;
[Errat-ibid {\bf 48} (1993) 1879];
A. Czarnecki and M. Skrzypek, Phys. Lett. B {\bf 449} (1999) 354.

\bibitem{C8}
T. Kinoshita and W.B. Lindquist. Phys. Rev. Lett. {\bf 47} (1981) 1573;
Phys. Rev D {\bf 271} (1983) 867;
ibid. {\bf 27} (1983) 877;
ibid. {\bf 27} (1983) 886;
ibid. {\bf 39} (1989) 2407;
ibid. {\bf 42} (1990) 636;
T. Kinoshita, B. Nizic and Y. Okamoto, ibid. {\bf 41} (1990) 593;
ibid. {\bf 52} (1984) 717.

\bibitem{C10}
  T. Aoyama, {\it et al.}, 
  Phys.\ Rev.\  D {\bf 78} (2008) 113006;
  ibid.  {\bf 78} (2008) 053005.

\bibitem{AGdR08}
  J.P. Aguilar, D. Greynat and E. de Rafael,
  Phys.\ Rev.\  D {\bf 77} (2008) 093010.

\bibitem{DH98}
 M. Davier and A. H\"ocker,
  Phys.\ Lett.\  B {\bf 435} (1998) 427.

\bibitem{KRA97}
B. Krause,  Phys.\ Lett.\  B {\bf 390} (1997) 392.

\bibitem{PdRV09}
J. Prades, E. de Rafael and A. Vainshtein,  arXiv:0901.0306
in ``Lepton Dipole Moments'', B.L. Roberts and W.J. Marciano (eds)
  (World Scientific, Singapore, 2009),
Advanced Series on Directions in High Energy Physics, Vol. 20.


\bibitem{CKM96}
 A. Czarnecki, B. Krause and W.J. Marciano,
  Phys.\ Rev.\ Lett.\  {\bf 76} (1996) 3267.

\bibitem{EWone}
R. Jackiw and S. Weinberg, Phys. Rev. D {\bf 5} (1972) 2396;
I. Bars and M. Yoshimura, ibid. {\bf 6} (1972) 374;
K. Fujikawa, B.W. Lee and A.I. Sanda, ibid.  {\bf 6} (1972) 2923;
G. Altarelli, N. Cabibbo and L. Maiani, Phys. Lett. B {\bf 40} (1972) 415;
W.A. Bardeen, R. Gastmans and B. Lautrup, 
Nucl. Phys. B {\bf 46} (1972) 319.

\bibitem{KKS92}
T.V. Kukhto {\it et al}, Nucl. Phys. B {\bf 371} (1992) 567.

\bibitem{EWtwo}
A. Czarnecki, B. Krause and W.J. Marciano,
Phys. Rev. D  {\bf 52} (1995) 2619;
T. Kaneko and N. Nakazawa,  arXiv:hep-ph/9505278;
S. Peris, M. Perrottet and E. de Rafael, Phys. Lett. B
{\bf 355} (1995) 523;
G. Degrassi and G.F. Giudice, Phys. Rev. D {\bf 58} (1998) 053007;
M. Knecht {\it et al}, JHEP {\bf 11} (2002) 003;
A. Czarnecki, W.J. Marciano and A. Vainshtein,
Phys. Rev D {\bf 67} (2003) 073006;
Acta. Phys. Polon. B {\bf 34} (2003) 5669;
S. Heinemeyer, D. St\"ockinger and G. Weiglein, 
Nucl. Phys. B {\bf 699} (2004) 103.

\bibitem{LOhad}
C. Bouchiat and L. Michel, J. Phys. Radium {\bf 22} (1961) 121;
L. Durand, Phys. Rev. {\bf 128} (1962) 441;
[Erratum-ibid {\bf 129} (1963) 2835];
S.J. Brodsky and E. de Rafael, Phys. Rev. {\bf 168} (1968) 1620;
M. Gourdin and E. de Rafael, Nucl. Phys. B {\bf 10} (1969) 667.

\bibitem{CMD207}
R.R. Akhmetshin {\it et al}, [CMD2 Collaboration],
Phys. Lett. B {\bf 648} (2007) 28;
Phys. Lett. B {\bf 466} (1999) 392;
V.M. Aulchenko {\it et al}, [CMD2 Collaboration],
JETP Lett. {\bf 84} (2006) 413;
ibid. {\bf 82} (2005) 743.

\bibitem{CMD206}
I.B. Logashenko, Nucl. Phys. B (Proc. Suppl.) {\bf 162} (2006) 13.

\bibitem{SND06}
  M.N. Achasov {\it et al.},
  Nucl.\ Phys.\ B (Proc.\ Suppl.)  {\bf 162} (2006) 11;
JETP Lett. {\bf 103} (2006) 103.

\bibitem{KLOE09}
F. Ambrosino {\it et al} [KLOE Collaboration],
Phys. Lett. B {\bf 670} (2009) 285.

\bibitem{BABAR09}
B. Aubert  {\it et al} [BABAR Collaboration],
Phys. Rev. Lett.  {\bf 103} (2009) 231801.

\bibitem{DAV09b}
M. Davier {\it et al}, 
  arXiv:0908.4300;
       M. Davier  {\it et al}, [BABAR Collaboration],
  Nucl.\ Phys. B (Proc.\ Suppl.)  {\bf 189} (2009) 222.

\bibitem{DAV09a}
 M. Davier {\it et al.},
  arXiv:0906.5443.

\bibitem{HAG07}
K. Hagiwara, {\it et al}, Phys. Lett. B {\bf 649} (2007) 173. 

\bibitem{HK98} 
M. Hayakawa and T. Kinoshita,  Phys.\ Rev.\ D {\bf 57 } (1998) 465
  [Erratum-ibid.\ D {\bf 66} (2002) 019902];
M. Hayakawa, T. Kinoshita and A.I. Sanda,
 Phys.\ Rev.\ Lett.\  {\bf 75} (1995) 790.

\bibitem{BPP96} 
J. Bijnens, E. Pallante and J. Prades,
  Nucl.\ Phys.\ B {\bf 474} (1996) 379;
Phys.\ Rev.\ Lett.\  {\bf 75} (1995) 1447
  [Erratum-ibid.\  {\bf 75} (1995) 3781];
  Nucl.\ Phys.\ B {\bf 626} (2002) 410.

\bibitem{BP01}
J. Bijnens and F. Persson, Lund University Master Thesis,
hep-ph/0106130.

\bibitem{KN02}  M. Knecht and A. Nyffeler,
  Phys.\ Rev.\ D {\bf 65} (2002) 073034;
M. Knecht {\it et al},  Phys.\ Rev.\ Lett.\  {\bf 88} (2002) 071802.

\bibitem{MV04} K. Melnikov and A. Vainshtein,
  Phys.\ Rev.\ D {\bf 70} (2004) 113006.

\bibitem{DB08}
 A.E. Dorokhov and W. Broniowski,  
Phys.\ Rev.\  D \textbf{78} (2008) 073011.

\bibitem{NYF09}
A. Nyffeler, Phys. Rev. D \textbf{79} (2009) 073012

\bibitem{LATT09}
T. Blum and S. Chowdhury, Nucl. Phys. B (Proc. Suppl.)
{\bf 189} (2009) 251;
M. Hayakawa, {\it et al},   PoS {\bf LAT2005} (2006) 353.

\bibitem{reviews}
  J. Prades,  PoS {\bf (CD09)} 079;
  arXiv:0907.2938;
  arXiv:0905.3164 (to be published in Eur. Phys. J C);
 Nucl.\ Phys. B (Proc.\ Suppl.)  {\bf 181-182} (2008) 15;
 J. Bijnens and J. Prades,
  Mod.\ Phys.\ Lett.\  A {\bf 22} (2007) 767;
  Acta Phys.\ Polon.\  B {\bf 38} (2007) 2819.

\bibitem{PRA01}
 J. Prades,
  arXiv:hep-ph/0108192.

\end{thebibliography}
\end{document}